# A Three-Layer Framework for Measuring Names and Its Census Application on a Token Launchpad

Dingding Cao[a], Yujing Zhong[a], Han Wang[a], Xian Pan[b], Rizwan Akhtar[c], Wei Yang[a,*]

[a] *School of Big Data, Baoshan University, Baoshan 678000, Yunnan, China*
[b] *School of Business, Guangzhou College of Technology and Business, Guangzhou 510850, China*
[c] *School of Computing Sciences, Pak-Austria Fachhochschule: Institute of Applied Sciences and Technology (PAF-IAST), Haripur, Pakistan*
* Corresponding author. E-mail: bsxy10202@bsc.edu.cn

## Abstract

Evidence that asset names influence market behavior continues to accumulate, yet the measurement of names themselves lacks a standardized foundation. Existing operationalizations of processing fluency are built largely on alphabetic scripts and do not apply to Chinese names. The cultural references carried by names have relied on manual coding, which does not scale to large samples. Competitive relations among names, including name reuse and semantic crowding, have received almost no systematic measurement. This paper constructs a census-level dataset of all 513,647 naming attempts on the four.meme token launchpad on BNB Chain between February and June 2026 and proposes a three-layer measurement framework for names. The form layer measures the processing fluency of a name as a linguistic object, extending fluency measurement to Chinese corpora through 38 deterministic features adapted to the Chinese writing system. The reference layer measures the cultural and social referents of names, combining a six-dimension codebook, double-blind two-round human coding, and full-scale expansion by a large language model, with reliability reported on three benchmarks: human double-blind agreement, human–machine replication, and machine self-consistency. The relation layer measures the position of a name within the contemporaneous population of names, covering name lodes, semantic crowding, and lexical variation. The three layers are nearly orthogonal, with pairwise Spearman correlations no greater than 0.11, each carrying independent information. The census reveals four population-level regularities: the name space expands according to Heaps' law ($V = 1.24 \cdot N^{0.921}$, $R^2 = 0.999$ on logarithmic scales), with the share of new names declining monotonically from 52.0% to 37.8%; name reuse is heavy-tailed (tail exponents 2.19 to 2.67), with the largest same-name series reaching 1,130 tokens; the median interval between successive uses of the same name is under three minutes and decomposes by creator identity into two separable time scales, 0.7 minutes within creator and 5.6 minutes across creators; and cultural events trigger immediate quantity responses, with 4,128 new tokens created within six hours of the Spring Festival Gala broadcast, 12.3% of which used festival-related names. The measurement framework, codebook, annotated data, and all code are released with the paper, and the framework is largely portable to other naming settings such as app stores, domain names, and social media hashtags.

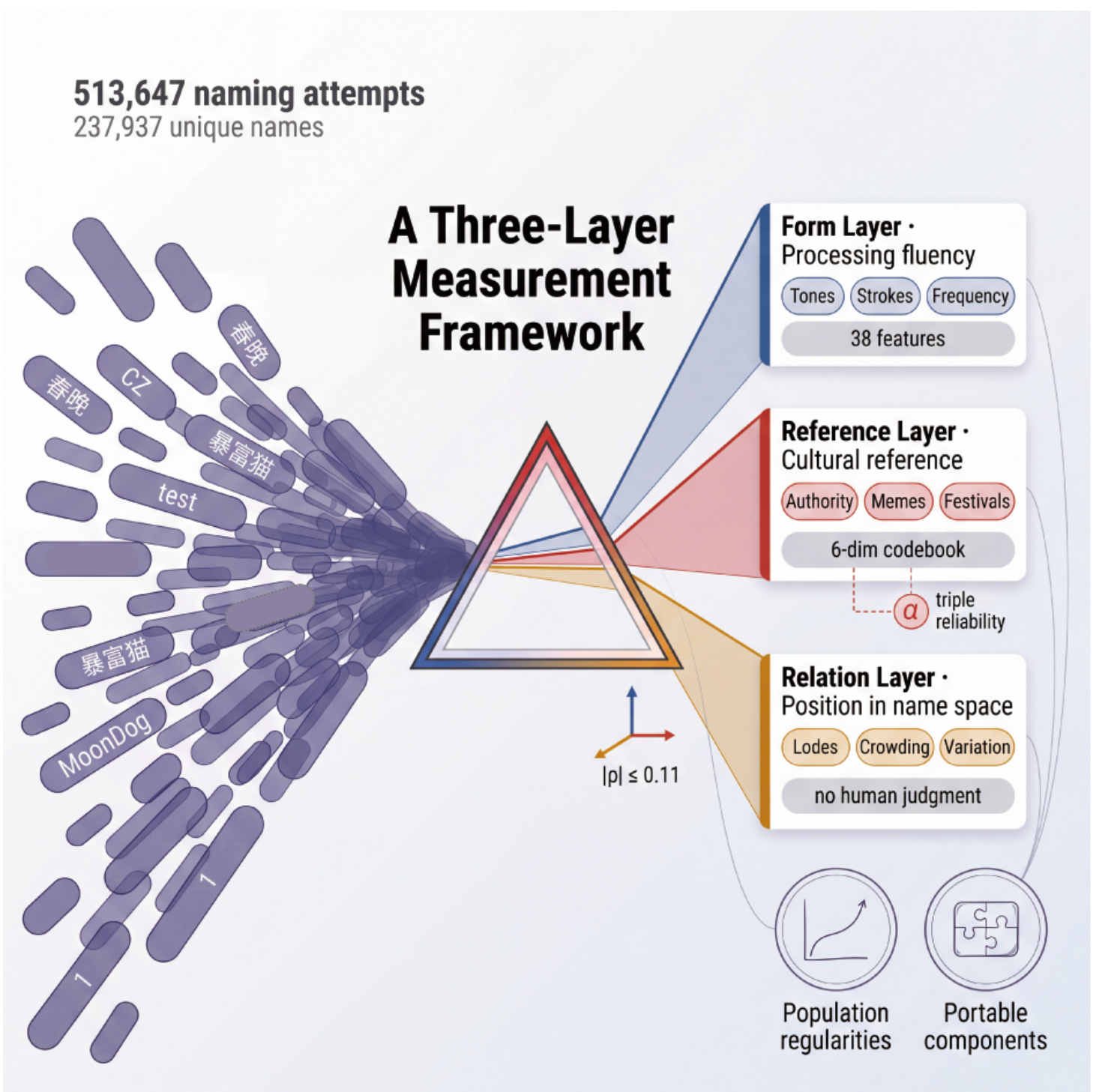

## Highlights

- A census of all 513,647 naming attempts, including tokens that never traded.
- A three-layer framework extending fluency measurement to Chinese names.
- A triple-benchmark reliability protocol for LLM-assisted content coding.

## 1. Introduction

Evidence that names influence economic behavior spans multiple markets. In equity markets, newly listed companies with pronounceable tickers earn higher returns in the days following listing [1], companies with fluent names have broader shareholder bases and higher valuations [2], and a name change alone can move prices: firms that added ".com" to their names during 1998–1999 experienced significantly positive market reactions [3], while firms that removed ".com" after the collapse of the internet bubble also experienced positive reactions [4], the two together indicating that the market responds to the timely salience of a name rather than to its content. Comparable evidence appears in fund markets, where renamed funds attract greater inflows and the renaming effect concentrates among funds whose names change most substantially [5], and where funds whose names drift toward hot investment styles attract additional subscriptions [6]. In venture and IPO markets, pronounceable names reduce underpricing [7]; at the level of retail investors, name familiarity is a stable predictor of portfolio choice [8,9].

Further research shows that even text carrying no fundamental information can affect investor behavior [10]. The predictive power of name fluency for stock returns survives controls for standard risk factors and concentrates in small-capitalization, high-attention stocks [11]. This cross-sectional pattern, however, does not identify the source of the effect. Because small-capitalization stocks have thin information environments, high retail participation, and attention-driven trading, mechanisms based on information, preferences, and attention all generate empirical predictions consistent with the same pattern. Discriminating among the three requires releasing the concept of a "name" from a single fluency summary and measuring its information-related, preference-related, and attention-related attributes separately.

The psychological foundation of name-fluency effects is generally attributed to processing fluency theory: information that is easier to process is more readily judged true, familiar, and likable [12,13]. The effect extends beyond financial settings; for example, individuals with easily pronounced names advance faster in their careers [14,15], and food additives with pronounceable names are judged safer [16].

A second psychological thread, parallel to fluency, concerns attention. Investor attention is a scarce resource, and salient stimuli, such as news coverage, extreme returns, and recognizable names, compete for that resource and thereby affect trading and prices [17,18].

A third thread comes from coordination games. When an asset lacks fundamental value to anchor its price, the name, as a commonly visible salient attribute, can serve as a focal point on which market participants coordinate their expectations, thereby affecting equilibrium prices [19,20]. Standard information economics, however, predicts the opposite. Because naming a project is nearly costless and unverifiable after the fact, it constitutes classic cheap talk: unless sending a signal requires real cost [21], such signals carry no information when the sender has a stake in the outcome, and rational investors should ignore them [22,23]. This tension leaves the question of whether names affect prices theoretically open.

The conclusions of this literature jointly depend on name attributes being properly measured, and current measurement practice exhibits three systematic gaps. The first is a language gap. Existing measures of processing fluency are built on alphabetic scripts, typically operationalized as the pronounceability of letter sequences and bigram frequency [1,2]. Chinese characters are not spelled from letters, so these measures do not transfer. For Chinese, the processing load of a name is jointly determined by several dimensions that are mutually independent and must be measured separately: visual complexity at the character level, with recognition slowing as stroke counts rise [24,25]; phonological structure, including tones and syllable composition [26]; character frequency [27,28]; and lexical status, that is, whether the name constitutes a real word or an idiom [29,30]. Consequently, although Chinese-named assets account for a substantial share of global crypto-asset issuance, they fall almost entirely outside the reach of existing measurement tools, and Chinese corpora are nearly absent from this research stream.

The second is a scale gap. The cultural references carried by names, such as authority borrowing, puns, and festival vocabulary, have historically been identified through manual coding. The methodological canon of manual coding is itself designed for small samples [31], and sample sizes in existing studies are typically in the hundreds, far short of modern issuance populations that run to the hundreds of thousands. Large language models make large-scale content coding feasible [32,33], but reliability norms for their use as coding instruments are not yet established: most applications report only model–human agreement, not the stability of the model under repeated coding, and few specify the adjudication and disposition rules for each dimension before coding begins.

The third is a dimensionality gap. Existing measures treat each name as an isolated object, yet part of the economic meaning of a name derives not from the name itself but from its position within the population of all existing names: how many times the name has been used before, and how many semantically similar competitors it faces at the same time. The tools for characterizing this relational dimension have long been mature in linguistics and information retrieval, including Heaps' law for vocabulary growth [34,35], Zipf's law for frequency distributions [36], statistical estimation methods for heavy-tailed distributions [37], and models of name-popularity dynamics from cultural evolution [38,39], but these tools have not been widely introduced into research on asset naming.

This paper addresses the three gaps simultaneously in a setting exceptionally favorable to name measurement: the four.meme token launchpad on BNB Chain [40], where any address can issue a token for a fee of roughly USD 0.05, making issuance costs nearly zero. At issuance the creator sets four principal attributes: the token name, the ticker symbol, a text description, and an avatar image. The name is the primary identifier of the token in trading interfaces and aggregator listings; the symbol largely coincides with or abbreviates the name, so the two carry highly overlapping information. The description appears only in a secondary position on the token detail page and receives limited attention in a trading environment dominated by rapid browsing. The avatar image, although also optional, engages processing mechanisms fundamentally different from those of text, and placing images and text in the same measurement framework would introduce incommensurable dimensions. This paper therefore restricts its object to textual attributes, with the name at the core, leaving image attributes to future research.

Launchpads have been the dominant mechanism for on-chain token issuance since 2024, and parallel research on pump.fun on the Solana chain provides a point of comparison [41]. In terms of broader context, the systematic characteristics of crypto-asset markets are well documented [42,43]: price formation is retail-dominated and attention-driven [44,45], and manipulation and wash trading are widespread [46,47]. For present purposes the value of this setting lies in three features. Naming occurs continuously at very high frequency, yielding a sample orders of magnitude larger than in existing name research. Chinese names account for a substantial share, providing a natural corpus for measurement across writing systems. Most importantly, the name is fixed once and irrevocably before the token has any trading or price, issuance is nearly free, and no uniqueness constraint applies, so naming is a pure text decision that strictly precedes market outcomes and is nearly unconstrained by fundamentals.

For measurement research, the setting combines three conditions that are difficult to obtain jointly in traditional markets. First, the population of naming behavior is fully observable. Token issuance is completely recorded by on-chain factory-contract events, and every naming attempt is preserved regardless of subsequent success, including tokens that never traded, so measurement passes through no survivorship filter. This contrasts with equity or fund research, where only names that cross listing or registration thresholds enter the researcher's field of view and rejected naming attempts are unobservable. Second, name reuse arises naturally. Because the platform imposes no uniqueness constraint, the same name is registered repeatedly at different times by different addresses, providing direct observations for measuring the relational dimension; in settings with enforced uniqueness, such as stock tickers and domain names, such observations are impossible in principle. Third, the decision environment is highly homogeneous. All tokens share the same contract code, supply schedule, and pricing mechanism, so the name is one of the few attributes on which tokens substantively differ, and name differences are not confounded with governance structure, underwriting quality, or other attributes.

Around these issues this paper develops a systematic measurement framework for asset names, quantifying names along three mutually orthogonal dimensions: processing fluency, cultural reference, and inter-name relations. It extends fluency measurement to Chinese-named assets for the first time through linguistically grounded indicators adapted to Chinese characters, scales cultural content analysis to census size through large-language-model coding subjected to strict reliability testing, and characterizes several statistical regularities of name reuse and name-space expansion on a fully observed naming population. The framework, together with its reliability protocol and population benchmarks, is transferable to other markets and corpora, providing a measurement foundation for subsequent causal tests of names and market outcomes.

## 2. Data and Methods

### 2.1 Institutional setting

The research setting is the four.meme token launchpad on BNB Chain (BSC). Upon submission of a name and symbol and payment of an on-chain fee amounting to single-digit US dollars, the factory contract deploys the token instantly: a fixed supply of one billion units attached to a deterministic bonding curve on which trading begins. When funds raised on the internal market reach a threshold of 17.64 BNB (net of a 2% fee), liquidity migrates to a public decentralized exchange and the internal market closes. The threshold is a hard constant throughout the research window, with every quantile equal to the same value. Issuance involves no review, no queuing, and no disclosure obligations.

Under this regime, names possess four properties critical for measurement: they are free (adding nothing to deployment cost), unrestricted (any string can be registered), unverifiable (constituting no factual commitment), and unconstrained by uniqueness (names carry no trademark status, and the same name can be registered repeatedly by any address at any time). The fourth property matters most for measurement research: it makes the recurrence of identical name content at different times and in different circumstances a natural phenomenon, rendering name content and name circumstance separable in the data, which is what makes the relational dimension definable. Within the research window, 237,937 unique names correspond to 513,647 naming attempts; 71,585 names were used more than once, and the maximum reuse count for a single name reached 1,130. This paper refers to the set of tokens sharing the same normalized name as a name lode.

### 2.2 Cohort and data sources

The cohort was fixed in advance by calendar window: all tokens created by the four.meme factory contract between February 1 and June 30, 2026, with the window determined before any data collection. The cohort comprises 513,647 tokens (97,436; 176,907; 107,815; 81,338; and 50,151 by month), with no duplicate contract addresses, enumerated completely from factory creation events; 58,816 tokens (11.4%) never produced a single trade. Zero-trade tokens are a substantive component of the population rather than noise: their naming behavior arises from the same decision process as that of traded tokens, and excluding them would impose survivorship bias on any statement about the naming population. They are therefore retained throughout and reported separately.

Data come from a two-source pipeline. The discovery layer derives from an on-chain event index (Dune) [48]: creation events yield the complete token list, creator addresses, and original names; liquidity events yield migration; internal-market buy and sell events yield aggregates of early fund flows; and native transfers yield the source of each creator's first funding. The attribute and trade layer derives from OKX OnchainOS [49]: names and symbols, market snapshots, holder lists (top 100), and trade-level records (direction, price, USD amount, trader address, transaction hash) totaling 52,066,542 trades, covering the internal market and indexed decentralized-exchange venues, with each token crawled back to its creation time. The two sources are cross-checked at the token-enumeration layer. The name field is taken from on-chain creation events, which can be neither tampered with nor edited after the fact; this is the data-level reliability foundation of the object this paper measures.

### 2.3 Census completeness audit

The designation "census" is established by a completeness audit, not by sample size. The audit proceeds on three fronts (Table 1). With respect to field coverage, creator address, name, migration flag, internal-market aggregates, and trade counts are covered at 100%; holder lists at 97.09%; and market snapshots at 87.31%. Examination of the missingness structure shows that missing market snapshots are not random: within the missing group, the zero-trade share is 90.2%, the migration rate is 0%, and the mean number of trades is 0.11. Missingness concentrates almost entirely on tokens that never generated any market activity; that is, it is defined by the very fact of “no trading” and constitutes structural non-measurement rather than data loss. Finally, this non-measurement does not touch the naming side: the name field of zero-activity tokens is fully measurable and retained throughout, so the structural gaps in market fields leave the completeness of every naming-side measure unaffected.

**Table 1  Census completeness audit**

**Panel A: Field coverage**

| Field | Source | Coverage |
|---|---|---|
| Creator address | On-chain creation events | 100.00% |
| Name | On-chain creation events | 100.00% |
| Migration flag | On-chain liquidity events | 100.00% |
| Early internal-market inflows (aggregate) | On-chain buy/sell events | 100.00% |
| Trade count | OKX trade index | 100.00% |
| Source of creator’s first funding | Native transfers | 99.36% |
| Holder list (top 100) | OKX holder API | 97.09% |
| Market snapshot (market cap/price) | OKX market API | 87.31% |

**Panel B: Structure of missing market snapshots**

| | Tokens | Zero-trade share | Migration rate | Mean trades |
|---|---|---|---|---|
| With market snapshot | 448,451 | 0.00% | 0.55% | 116.09 |
| Without market snapshot | 65,196 | 90.21% | 0.00% | 0.11 |

Notes: Panel B shows that missing market snapshots concentrate almost entirely on tokens that never generated market activity, constituting structural non-measurement rather than random missingness.

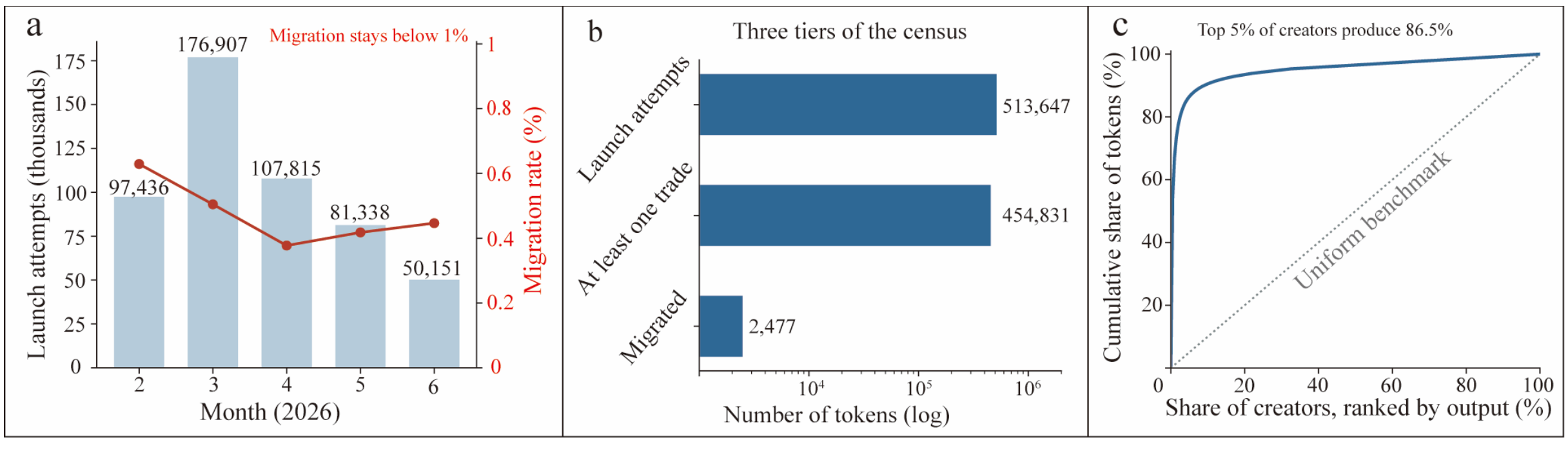

Figure 1. Market and census overview. (a) Monthly issuance attempts (bars, left axis) and migration rate (line, right axis); (b) token counts at three tiers: issuance attempts, at least one trade, and migration (logarithmic horizontal axis); (c) cumulative share of tokens by creators ranked in descending order of output, with the dotted line as the uniform-issuance reference.

The composition of the corpus is as follows. A total of 511,664 tokens carry non-empty names (1,983 empty names are listed separately), corresponding to 237,937 unique names: Chinese 41.7%, English 44.3%, Chinese–English mixed 11.8%, and other 2.3%. To our knowledge this is the first large-scale corpus in the name-research literature in which Chinese is a principal component. Creators number 35,677 addresses, of which 32.1% issued two or more tokens; prolific addresses contribute 95.3% of all tokens, and the most prolific address issued 9,750 tokens in five months.

### 2.4 The three-layer measurement framework

The three layers are distinguished by three mutually independent ways in which a name carries information. The form layer measures the processing load of the name as a linguistic object in itself, that is, whether a name is easy to read, remember, and say, without regard to what it points to. The reference layer measures what objects outside the corpus the name points to, such as authoritative institutions, cultural symbols, or seasonal events, without regard to its linguistic form. The relation layer measures the position of the name within the population of names at the moment it appears, including how many times it has been used before and how many semantic neighbors surround it, independent of both content and form. The measurement system follows three design principles. Every field that can be computed deterministically is computed deterministically, and every field requiring human judgment is constrained by a codebook with reliability reported, ensuring reproducibility. All measurement is completed before any trading-outcome field is connected, after which no parameter is adjusted, ensuring that measurement precedes outcomes. Chinese and English each receive their own adapted operationalizations, with neither bent to fit the other, ensuring linguistic parity. All 237,937 unique names enter measurement.

#### *2.4.1 The form layer: fluency measures adapted to the Chinese writing system*

Existing measures of processing fluency are built largely on alphabetic scripts, represented by the pronounceability of letter sequences and bigram frequency, and cannot be applied directly to Chinese names. Chinese characters are not spelled from letters; their processing load arises from roughly four mutually independent sources: stroke complexity at the visual level, tone and syllable structure at the phonological level, familiarity as reflected in character frequency, and lexical status, that is, whether the name constitutes a real word or an idiom.

On this basis the paper constructs 20 fields for Chinese names, covering the phonological, graphemic, and lexical facets. English names receive 12 parallel fields constructed on the principle of same structure, different sources. A further 6 cross-language fields record the orthographic composition of the name. The form layer thus comprises 38 fields in total; Table 2 reports the definitions and distributions of all fields by group. Names containing at least one

Chinese character enter the Chinese fields, names containing at least one Latin letter enter the English fields, and mixed-script names accordingly enter both sets.

All fields are computed deterministically with version-pinned Python: segmentation and dictionary lookup use jieba (498,113 entries), pinyin and tones use pypinyin, character and word frequencies use wordfreq, and stroke counts use strokes. A reference table for the 4,819 Chinese characters appearing in the corpus is released with the paper; stroke counts have no gaps, and only 53 characters lack frequency records. Given the name string and resource versions, any researcher recomputing the fields reproduces them value by value. Notably, the only genuine ambiguity among the 38 fields concerns tone: 30.6% of Chinese characters are heteronyms with ambiguous tone. The paper adopts whole-name contextual reading as the primary basis and compares it with character-by-character reading; only 6.65% of Chinese names differ in tone entropy between the two bases, with a mean difference of 0.117, which is the measured upper bound of measurement error for this dimension. The form layer can therefore serve as the measurement-error benchmark of the entire framework: any argument that invokes measurement attenuation to challenge downstream conclusions must first fail on the form layer.

The correlation structure among dimensions supports this design. For the six core Chinese dimensions, the mean absolute pairwise Spearman correlation is 0.098 with a maximum of 0.374, and the first principal component explains only 25.2% of total variance; the English-side counterparts are 0.277, 0.544, and 37.7%. The dimensions are neither redundant nor unrelated, retaining moderate association through the common construct. Because Chinese assigns one syllable per character, syllable count and character count are numerically identical for purely Chinese names and should not enter a model simultaneously with name length.

**Table 2 Distributions of all 38 form-layer fields (unit of observation: unique names)**

**Panel A: Chinese fields (20; n = 136,498 names containing Chinese characters)**

| Field | Variable | Mean | Median | P10–P90 |
|---|---|---|---|---|
| Character count | zh_n_char | 5.95 | 5.00 | 2.00–11.00 |
| Syllable count | zh_n_syllables | 5.95 | 5.00 | 2.00–11.00 |
| Tone entropy (contextual) | zh_tone_entropy | 0.625 | 0.597 | 0.406–1.000 |
| Distinct tones | zh_n_distinct_tones | 3.15 | 3.00 | 2.00–5.00 |
| Tone entropy (per character) | zh_tone_entropy_naive | 0.624 | 0.597 | 0.406–1.000 |
| Reduplication flag † | zh_has_reduplication | 4.4% | — | — |
| Alliteration (shared onset share) | zh_alliteration | 0.124 | 0.000 | 0.000–0.333 |
| Rhyme (shared rime share) | zh_rhyme | 0.130 | 0.100 | 0.000–0.333 |
| Mean strokes | zh_strokes_mean | 7.55 | 7.44 | 5.50–9.67 |
| Maximum strokes | zh_strokes_max | 11.52 | 12.00 | 8.00–15.00 |
| Total strokes | zh_strokes_total | 44.57 | 39.00 | 17.00–80.00 |
| Mean character-frequency quantile | zh_charfreq_mean | 5.10 | 5.09 | 4.43–5.80 |
| Minimum character-frequency quantile | zh_charfreq_min | 4.10 | 4.16 | 3.25–4.84 |
| Out-of-vocabulary characters | zh_n_oov_char | 0.001 | 0.00 | 0.00–0.00 |
| Heteronym characters | zh_n_heteronym_char | 2.85 | 2.00 | 1.00–6.00 |
| Segmented tokens | zh_n_tokens | 3.31 | 3.00 | 1.00–6.00 |
| Wordhood | zh_wordhood | 0.654 | 0.667 | 0.000–1.000 |
| Single-word flag † | zh_is_single_word | 11.2% | — | — |
| Idiomaticity flag † | zh_idiomaticity | 1.2% | — | — |
| Whole-name frequency quantile ‡ | zh_wholename_zipf | 3.42 | 3.43 | 1.83–4.96 |

**Panel B: English fields (12; n = 133,336 names containing Latin letters)**

| Field | Variable | Mean | Median | P10–P90 |
|---|---|---|---|---|
| Token count | en_n_token | 1.83 | 2.00 | 1.00–3.00 |
| Letter count | en_n_letter | 8.38 | 8.00 | 3.00–15.00 |
| Syllable count | en_n_syllables | 2.81 | 3.00 | 1.00–5.00 |
| Letter-bigram log probability | en_bigram_lp | −1.243 | −1.184 | −1.546 – −1.019 |
| Letters per token | en_letters_per_token | 4.81 | 4.33 | 2.00–8.00 |
| Consonant-cluster density | en_consonant_cluster | 0.089 | 0.000 | 0.000–0.375 |
| Capitalization share | en_caps_ratio | 0.307 | 0.200 | 0.000–1.000 |
| Mean word-frequency quantile | en_wordfreq_mean | 4.19 | 4.23 | 2.74–5.61 |
| Minimum word-frequency quantile | en_wordfreq_min | 3.68 | 3.77 | 2.19–5.10 |
| Out-of-vocabulary tokens | en_n_oov_token | 0.25 | 0.00 | 0.00–1.00 |
| Wordhood | en_wordhood | 0.792 | 1.000 | 0.000–1.000 |
| Formulaicity | en_formulaicity | 1.65 | 0.00 | 0.00–4.34 |

**Panel C: Cross-language fields (6; n = 237,937 all names)**

| Field | Variable | Mean | Median | P10–P90 |
|---|---|---|---|---|
| CJK character count | n_cjk | 3.42 | 3.00 | 0.00–9.00 |
| Latin letter count | n_latin | 4.70 | 3.00 | 0.00–13.00 |
| Mixing ratio | mix_ratio | 0.047 | 0.000 | 0.000–0.263 |
| Digit count | n_digit | 0.15 | 0.00 | 0.00–0.00 |
| Contains-digit flag † | has_digit | 6.8% | — | — |
| Length-censoring flag † | len_censored | 3.1% | — | — |

Notes: † denotes binary flags, for which the positive share is reported. ‡ The whole-name frequency quantile is defined only for names constituting a single token (n = 21,627). Character and word frequencies are Zipf quantiles (1–8; larger values indicate more common). Mixed-script names (34,223) enter both Panels A and B. Tone entropy is normalized entropy (0 for all-identical tones, 1 for uniform tones); wordhood is the share of characters (letters) covered by multi-

character dictionary words; idiomaticity requires four characters forming a single dictionary entry; formulaicity is the phrase-level frequency quantile of a multi-word name taken as a whole.

### *2.4.2 The reference layer: codebook, double-blind coding, and LLM expansion*

The reference layer identifies what cultural and social objects a name points to outside the corpus, with categories in two groups. Mechanically decidable categories, such as containing digits, matching an event word list, or being a homophone-collision candidate, are implemented by rules and carry no judgment error. Categories requiring judgment comprise six dimensions. Authority borrowing (D1) denotes names that appropriate the symbolic capital of established authorities such as exchanges, public chains, and well-known figures, with a homophone-variant subdimension (D1b); ironic self-reference (D2) denotes confessional naming that names the scam as such; cultural memes (D3) denotes invocation of internet catchphrases and cultural symbols; seasonal events (D4) denotes reference to festivals and current events; financialized blessings (D5) denotes benediction terms such as “get rich” and “fortune” used as asset names; and non-substantive naming (D6) denotes pure gibberish or placeholders. Rhetorical devices specific to Chinese, such as homophone substitution, digit culture, and festival vocabulary, are carried by these dimensions. The codebook specifies decision rules, positive and negative examples, and boundary cases for each dimension.

Human coding first establishes the gold standard. Two coders independently and blindly coded a stratified sample of 408 names, seeing only the names themselves in randomized order with no access to any trading data; reliability is reported as Krippendorff’s α [31] over two rounds. The first round fell short on several dimensions, with D2 at 0.248 and D5 at 0.313. After a codebook alignment procedure that added five boundary rules (the codebook’s decision rules themselves unchanged), a full independent second-round recoding raised α across the six dimensions to between 0.756 and 0.982, with D1b reaching 1.000. Both rounds are reported side by side rather than the second alone, because an α obtained after supplementing rules with knowledge of where agreement failed is no longer an unbiased estimate of independent agreement, and reporting only the second round would overstate reliability.

Full-scale expansion is performed by a large language model. The model is DeepSeek-V4-Flash-0731, called through the official API with temperature 0, a fixed random seed of 42, and batches of 40 names. The actual model version returned by the server was logged batch by batch, and no version drift occurred. To rule out within-batch misalignment, the model was required to echo each name verbatim for comparison, with mismatches retried and bisected down to single names. Expansion followed a four-stage protocol: the gold standard was randomly split in half into a calibration set (202 names) and a validation set (206 names); prompts were iterated only on the calibration set; the same prompt was then run twice independently on the same names to measure machine self-consistency; a single run on the validation set then yielded human–machine reliability uncontaminated by prompt tuning, with the validation set permitted exactly one use; upon passing validation, coding was expanded to all 237,937 unique names.

Reliability therefore has three benchmarks (Table 3). Human double-blind α measures the reliability of the gold standard itself; human–machine α measures the model's replication of that standard; and machine self-consistency α measures the agreement of two runs under the same prompt, forming an upper bound on human–machine reliability, since a dimension on which the model is unstable with respect to itself cannot stably replicate the human standard.

The usage tier of each dimension is adjudicated by disposition rules written into the codebook in advance: a dimension qualifies for confirmatory use only if the lower bound of the 95% confidence interval of second-round human α is at least 0.667 and machine self-consistency α is at least 0.7. The alignment procedure permits supplementing boundary rules and one second-round recoding but no third round; dimensions still failing the bar are demoted to exploratory without exception and may not be rescued by repeated recoding until they pass. The codebook and all decision rules are released with the paper.

Accordingly, D1, D3, and D5 are confirmatory; D2 is demoted to exploratory because machine self-consistency is only 0.655; D4 and D6 are demoted because the lower bounds of their human confidence intervals fall short. The disagreements on D4 are not random but directional: the two coders disagreed on 14 names, 12 of them in the same direction, one coder judging an event reference present and the other absent, concentrated in names such as "New Year dumplings" in which festival words serve only as modifiers without pointing to a specific event. Such boundary cases are not covered by the codebook; under the rules above no third round is conducted, and D4 is demoted to exploratory.

**Table 3 Reliability of the judgment dimensions in the reference layer**

| Dimension | Human α (Round 1) | Human α (Round 2) | Round-2 95% CI | Human–machine α | Machine self-consistency α | Full-scale positive rate | Usage tier |
|---|---|---|---|---|---|---|---|
| Authority borrowing (D1) | 0.726 | 0.960 | [0.930, 0.985] | 0.877 | 0.979 | 23.94% | Confirmatory |
| Ironic self-reference (D2) | 0.248 | 0.982 | [0.937, 1.000] | 0.648 | 0.655 | 0.55% | Exploratory |
| Cultural memes (D3) | 0.561 | 0.877 | [0.806, 0.934] | 0.695 | 0.824 | 9.05% | Confirmatory |
| Seasonal events (D4) | 0.670 | 0.756 | [0.617, 0.867] | 0.842 | 0.969 | 1.84% | Exploratory |
| Financialized blessings (D5) | 0.313 | 0.906 | [0.825, 0.970] | 0.817 | 0.973 | 2.12% | Confirmatory |
| Non-substantive naming (D6) | 0.906 | 0.808 | [0.649, 0.923] | 0.768 | 0.945 | 3.51% | Exploratory |

Notes: The full-scale positive rate is the share of positives among the 511,664 tokens with non-empty names after LLM expansion.

*2.4.3 The relation layer: order, crowding, and variation*

The relation layer measures the position of a name within the name space at the moment it appears and contains four feature families. The first is lode membership and naming order. All tokens sharing the same normalized name key (whitespace removed, casefolded) constitute a lode, and each token records its within-lode order and the time elapsed since the previous token of the same name. After normalization, the tokens used 220,045 distinct names in total, each name together with all tokens bearing it forming one lode. Of these, 148,460 lodes contain a single token, meaning the name was used exactly once in the entire research window; the remaining 71,585 lodes contain two or more tokens, meaning the same name was registered again at a different time, often by a different creator. Roughly one in three names was reused, and the maximum reuse count for a single name reached 1,130.

The second is semantic-neighborhood crowding, which characterizes the contemporaneous competition a name faces at birth. For each token, two windows are taken, the 24 hours and the 7 days preceding creation, and a count is made of how many other tokens' names share content words with the focal name, content words being those that carry real meaning as opposed to high-frequency function words such as "coin". The count is weighted by inverse document frequency: the rarer a shared word, the higher its weight, so colliding on an obscure word contributes far more crowding than colliding on a word everyone uses. Chinese and English content words are produced by their respective tokenizers and word lists and are not directly comparable in count, so the two languages are standardized separately before entering the index.

The third is lexical variation, which captures a name's imitation and modification of precedents. For each name, among all names appearing earlier in the corpus, the most similar one is sought and termed the precursor anchor; if the edit similarity between the two names is at least 0.85, meaning one can be transformed into the other with only a few character insertions, deletions, or substitutions, the anchor is deemed to exist, and the distance between the new name and its anchor is recorded as the variation measure. Retrieval proceeds in two steps: candidates are first screened rapidly by overlap of character trigrams, and edit similarity is then computed exactly on the candidates. Of all names, 6.4% have a precursor anchor.

The fourth is launch density at the minute, hour, and day levels, recording the total issuance volume within the same minute, hour, and day as the token's creation, that is, the instantaneous crowding at the moment of naming.

All relation-layer features are computed solely from name strings and on-chain timestamps and involve no human judgment. The only approximation is the blocked retrieval used for lexical variation, validated against full pairwise comparison on a random sample of 3,000 name keys with a recall of 97.72%; the missed pairs all lie near the similarity threshold, and the misses bias imitation coverage toward the conservative side rather than the permissive one.

*2.4.4 Aggregation and reproducibility*

The three layers are each aggregated into a subindex by a covariance-inverse-weighted GLS procedure [50], with feature directions determined programmatically by the sign of rank correlation with the aggregation target, without manual adjustment. All measurement targets the name field. The symbol field coincides exactly with the name for 51.1% of tokens, and in the remaining cases is mostly an abbreviation or truncation of the name; the median symbol length is 4 characters versus 7 for names, so the symbol's information content is length-limited and highly overlapping with the name, and the three-layer measurement accordingly takes the name as its object. The census master table, the 408 double-blind annotations, the codebook, the implementation code for the 38 form-layer fields and all relation-layer features, and the scripts producing every statistic in this paper are released with the paper; raw trade-level data are provided under controlled access owing to volume.

## 3. Results

### 3.1 Structure of the measurement framework

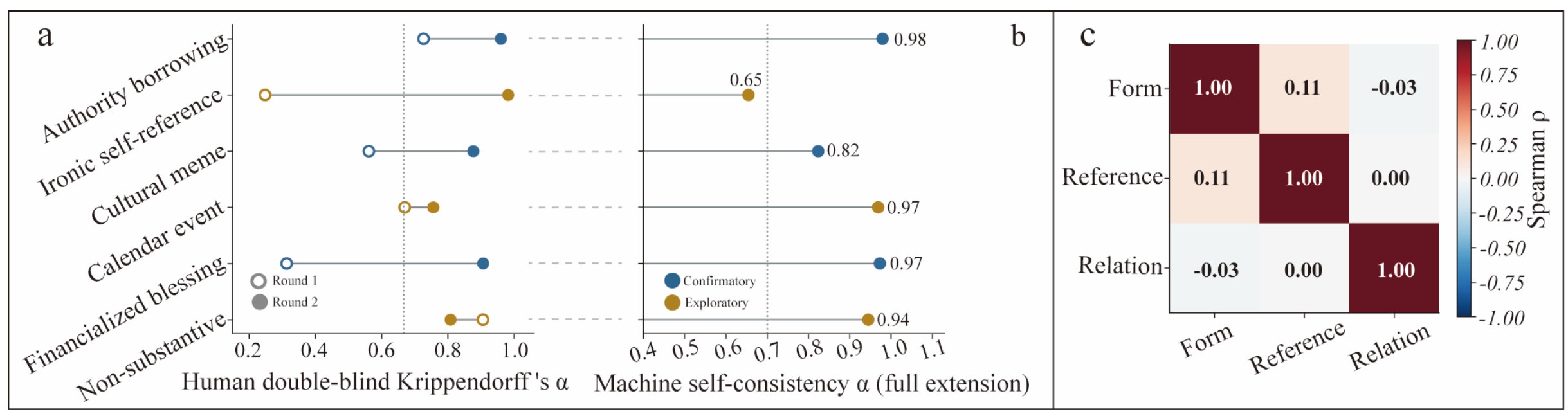

Figure 2. Reliability and structure of the three-layer measurement. (a) Human double-blind Krippendorff's α for the six judgment dimensions, open markers for Round 1 and filled markers for Round 2 (revised codebook), with the dashed line at the 0.667 threshold; (b) machine self-consistency α of the full-scale LLM expansion, with the dashed line at the 0.7 threshold and colors marking confirmatory versus exploratory use; (c) Spearman correlation matrix of the three subindices.

As shown in Figure 2, the pairwise Spearman correlations of the three subindices are all no greater than 0.11: 0.11 between the form and reference layers, −0.03 between the form and relation layers, and 0.00 between the reference and relation layers (Figure 2c). This nearly orthogonal correlation structure indicates that the three layers carry non-redundant information: whether a name is easy to process, what it refers to, and where it stands in the name space are three empirically independent dimensions. It follows that representing the overall attributes of a name by any single layer would discard all the variation contained in the other two.

The positive rates of the reference dimensions display a clear linguistic distribution (Table 4). The positive rate of authority borrowing (D1) among mixed Chinese–English names is 45.60%, far above 23.30% for purely Chinese and 19.63% for purely English names, consistent with the typical construction of mixed names, a Chinese modifier attached to an English brand word (for example "中国 Binance"), so that borrowing concentrates in this class. The positive rates of seasonal events (D4) and financialized blessings (D5) in Chinese names are three and four times those in English names (2.94% versus 0.84%, 3.86% versus 0.85%), consistent with the cultural specificity of these two devices rooted in the Chinese context; non-substantive naming (D6) is higher on the English side (3.18% versus 1.59%), as placeholders and gibberish more often use Latin characters. The differences in positive rates across language subgroups accord with the cultural provenance of each dimension's semantic content, indicating that the six dimensions possess discriminant validity for distinguishing naming content.

**Table 4 Positive rates by language subgroup and co-occurrence structure of the six reference dimensions**

| Dimension | Chinese (%) | English (%) | Mixed (%) | Full sample (%) | Highest co-occurrence lift |
|---|---|---|---|---|---|
| Authority borrowing (D1) | 23.30 | 19.63 | 45.60 | 23.94 | × D4 events: 2.06 |
| Ironic self-reference (D2) | 0.62 | 0.52 | 0.50 | 0.55 | × D3 memes: 4.21 |
| Cultural memes (D3) | 7.78 | 9.62 | 12.45 | 9.05 | × D2 irony: 4.21 |
| Seasonal events (D4) | 2.94 | 0.84 | 1.94 | 1.84 | × D5 blessings: 3.63 |
| Financialized blessings (D5) | 3.86 | 0.85 | 1.06 | 2.12 | × D4 events: 3.63 |
| Non-substantive naming (D6) | 1.59 | 3.18 | 2.51 | 3.51 | — |

Notes: Positive rates are within-subgroup shares after full-scale LLM expansion (Chinese n = 213,220; English n = 226,488; mixed n = 60,733 tokens). Co-occurrence lift = $P(A \wedge B)/[P(A) \cdot P(B)]$; values above 1 indicate that two dimensions co-occur on the same name more often than under independence. The distribution of the number of dimensions carried per name is 63.77% zero, 31.58% one, 4.52% two, and 0.13% three or more; most names deploy a single referential strategy, and stacking is rare.

The co-occurrence structure among dimensions provides a test from another angle. Define the co-occurrence lift of two dimensions as the ratio of their joint positive rate to the product of their individual positive rates; the ratio equals 1 if the two are independent, and values above 1 indicate that the two dimensions appear together on the same name more often than independence would imply. Each dimension was coded independently, and the coding process contained no information about inter-dimensional association, so the co-occurrence structure is a pattern that the labels exhibit spontaneously. The measured results accord with semantic expectation: the lift is 4.21 for ironic self-reference with cultural memes, 3.63 for seasonal events with financialized blessings, and 2.06 for authority borrowing with seasonal events, each pair carrying a direct semantic interpretation, as self-mocking names tend to speak through popular memes, festival names co-occur with words of blessing, and event-time borrowing concentrates on authority symbols. Semantically unrelated pairs, such as cultural memes with financialized blessings, show a lift of 0.46, below the independence benchmark. Co-occurrence rises and falls with semantic relatedness rather than spreading uniformly.

The nearly orthogonal correlation structure of the three layers, the linguistic differentiation of the reference dimensions, and the semantic coherence of the co-occurrence structure support the validity of the measurement framework from three mutually independent angles. The framework separates facets that ought to be conceptually independent and captures linguistic and cultural differences that genuinely exist in naming behavior.

### 3.2 The expansion law of the name space

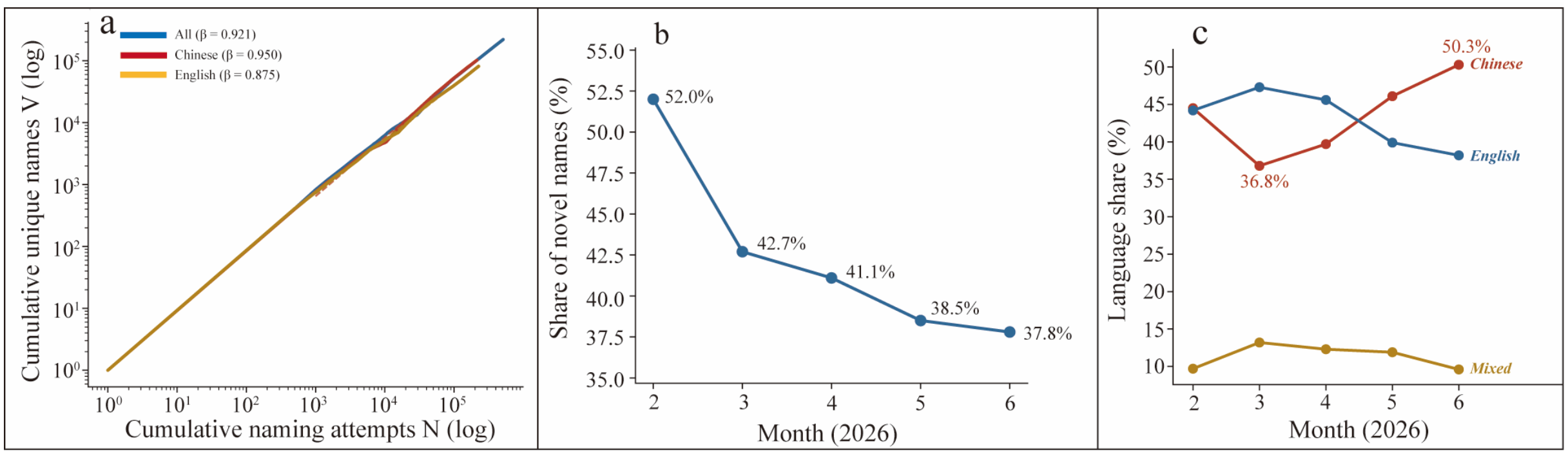

Figure 3. Dynamics of the name space. (a) Cumulative unique names against cumulative naming attempts (log–log), with the dashed line showing the Heaps'-law fit; (b) monthly share of new names; (c) monthly shares of the three language subgroups.

The expansion of the name space follows Heaps' law. Cumulative unique names V and cumulative naming attempts N satisfy $V = 1.24 \cdot N^{0.921}$, with $R^2 = 0.999$ on logarithmic scales (Figure 3a). An exponent below 1 implies diminishing marginal novelty of naming attempts: the probability that the N-th attempt produces a new name declines with N. The shape is consistent with vocabulary growth in natural-language corpora, where Heaps exponents typically lie between 0.4 and 0.6 [34]. The exponent in this market is markedly higher, near 0.92, indicating that namers are still to a large extent creating previously unseen strings and the name space is far from saturated. Language-specific estimates further show an exponent of 0.950 on the Chinese side and 0.875 on the English side; the higher novelty of Chinese naming is consistent with the linguistic fact that the combinatorial space of Chinese characters exceeds the reuse space of English vocabulary.

The share of new names declines monotonically from 52.0% in February to 37.8% in June (Figure 3b, Table 5), the direct monthly manifestation of an exponent below 1. Its rate of decline offers a yardstick comparable across markets: in a market where naming is costless and unconstrained by uniqueness, the crowding of the name space advances at roughly three percentage points per month.

The language composition is not stable but exhibits structural drift (Figure 3c, Table 5). The Chinese share falls from 44.5% in February to 36.8% in March and then climbs month by month to 50.3% in June, a swing of 13.5 percentage points; the English share falls from 47.3% to 38.2%. The drift is contemporaneous with the decline in total issuance, from 97,436 tokens in February to 50,151 in June: as the market cooled, the relative share of Chinese naming rose.

**Table 5 Monthly composition of the corpus and name-space expansion**

| Month (2026) | Issuance attempts | New names | New-name share (%) | Cumulative unique names | Chinese (%) | English (%) | Mixed (%) | Other (%) |
|---|---|---|---|---|---|---|---|---|
| February | 97,436 | 50,482 | 52.0 | 50,482 | 44.5 | 44.2 | 9.7 | 1.7 |
| March | 176,907 | 75,334 | 42.7 | 125,816 | 36.8 | 47.3 | 13.2 | 2.7 |
| April | 107,815 | 44,171 | 41.1 | 169,987 | 39.7 | 45.6 | 12.3 | 2.5 |
| May | 81,338 | 31,168 | 38.5 | 201,155 | 46.1 | 39.9 | 11.9 | 2.1 |
| June | 50,151 | 18,890 | 37.8 | 220,045 | 50.3 | 38.2 | 9.6 | 1.9 |

Notes: A new name is a name key not previously observed within the research window. Language shares are computed over tokens with non-empty names.

This drift bears directly on research design. Within the current window, Chinese names are more prevalent in the later, cooling phase of the market while English names concentrate in the earlier, heated phase, so the two language subgroups occupy periods with different market conditions, and any direct cross-language comparison would conflate language differences with period differences. Studies using this corpus for cross-language comparison must therefore incorporate time controls, stratifying by month or including period fixed effects. It should also be noted that this compositional drift becomes visible only under continuous observation covering all issuance; under outcome-based sampling, for example retaining only surviving or migrated tokens, monthly sample sizes would be insufficient to trace the month-by-month movement of shares.

### 3.3 The distribution of name reuse

Of the 220,045 lodes, 67.5% contain a single token, so most names are never used a second time; the size distribution of the remaining 32.5% is markedly heavy-tailed (Figure 4a, Table 6). Tail exponents estimated by the Hill estimator at three cutoffs lie between 2.19 and 2.67, stably within the interval from 2 to 3, implying that lode size has a finite mean but divergent variance: the average reuse count of a name is a stable statistic, while the dispersion of reuse counts is not. Any analysis using lode size as an explanatory variable must therefore employ logarithmic transformation or robust methods, as estimates on raw levels would be dominated by a handful of very large lodes.

Reuse concentrates heavily at the head (Table 6). The largest 1% of lodes contain 17.4% of all tokens, and the largest 10% contain 47.3%. Head lodes fall into two classes: generic placeholders, such as “test” and the digit “1”, and hot referent names, such as “CZ” and “Binance”, both instances of borrowing platform authority, corroborating the high positive rate of dimension D1 in the reference layer. The creator composition of the two classes differs starkly: the 1,130 tokens of “test” were contributed by 414 creators, while the 853 tokens of “SAT0 升级版” came from only 5. Reuse of the same magnitude can thus be generated by entirely different behaviors, and lode size alone does not characterize the reuse phenomenon.

Reuse is not only concentrated but rapid. The distribution of intervals between successive uses of the same name is markedly bimodal (Figure 4b), with a median under three minutes; a large share of copies occur within minutes of their predecessor, while another portion follows after more than a day. Naming activity is equally sensitive to external events: around the broadcast of the Spring Festival Gala, hourly creation counts and the share of festival-related names display two distinct temporal shapes (Figure 4c), the impact of the event on issuance volume being immediate.

**Table 6 Distributional statistics of lode size**

**Panel A: Quantiles and tail exponents**

| Statistic | Value | Statistic | Value |
|---|---|---|---|
| Share of single-use lodes | 67.5% | 90th percentile | 4 tokens |
| Number of reused lodes | 71,585 | 99th percentile | 20 tokens |
| Maximum lode size | 1,130 tokens | 99.9th percentile | 64 tokens |
| Hill tail exponent (xmin = 2) | 2.19 | Hill tail exponent (xmin = 5 / 10) | 2.43 / 2.67 |

**Panel B: Concentration and examples of head lodes**

| Concentration | Value | Head lode | Size and creators |
|---|---|---|---|
| Tokens in largest 1% of lodes | 17.4% | “test” | 1,130 tokens, 414 creators |
| Tokens in largest 5% of lodes | 36.0% | “1” | 630 tokens |
| Tokens in largest 10% of lodes | 47.3% | “CZ” / “Binance” | 476 / 445 tokens |
| | | “SAT0 升级版” | 853 tokens, 5 creators |

Notes: Lode size is the number of tokens under the same normalized name key. Hill tail exponents are estimated separately at three cutoffs (xmin = 2, 5, 10) to check robustness to cutoff choice; exponents between 2 and 3 indicate a finite mean with divergent variance. Concentration is the share of tokens contained in the largest lodes out of all 511,664 tokens with non-empty names.

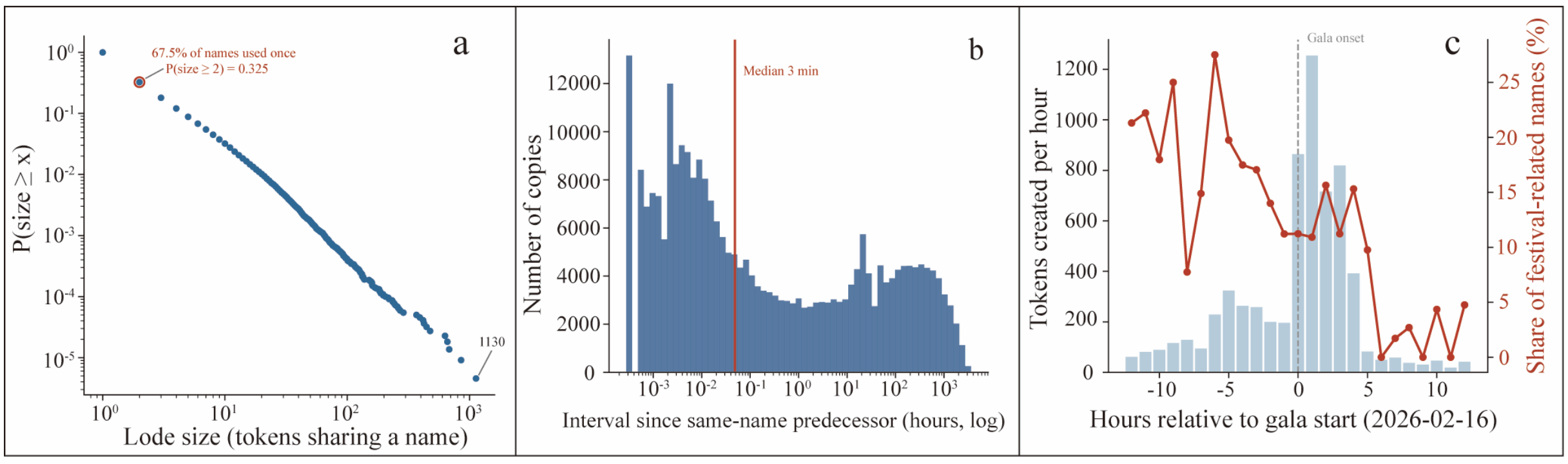

Figure 4. The naming ecology. (a) Survival function of lode size (log–log): the horizontal axis is lode size x and the vertical axis the share of lodes of size at least x; a power-law distribution appears as a straight line in these coordinates. The red circle marks P(size ≥ 2) = 0.325, that is, 67.5% of names are used exactly once, and the tail annotation marks the largest lode of 1,130 tokens. (b) Distribution of intervals between successive uses of the same name (logarithmic

horizontal axis), with a median under three minutes. (c) Hourly creation counts (bars, left axis) and the share of festival-related names (line, right axis) around the Spring Festival Gala broadcast (February 16, 2026, dashed line).

### 3.4 The two-scale temporal structure of copying

The bimodal interval distribution of same-name copying (Figure 4b) implies that copying mixes a fast and a slow process, and its decomposition requires knowing who executes each copy. Splitting by whether the copier and the predecessor's creator are the same address resolves the bimodality into two separable time scales (Figure 5a): the median interval is 0.7 minutes for same-creator copies (n = 46,616) and 5.6 minutes for cross-creator copies (n = 233,223), and the difference between the two distributions is highly significant under a Mann–Whitney test ($p < 0.001$). The two scales correspond to two behaviors: second-level same-creator repetition is batch deployment, a single operator redeploying same-name contracts within a very short window, while minute-to-day-level cross-creator repetition is imitative adoption of others' names. The two are indistinguishable in the name field itself but cleanly separable in the time and identity dimensions. Downstream research taking same-name tokens as its unit of analysis can therefore stratify along these lines; otherwise the capacity of a single operator and the imitation of many would be conflated into one phenomenon.

The composition at the lode level corroborates the same stratification (Figure 5b, Table 7). Of the 71,585 reused lodes, those produced by a single creator account for 9.2% and contain 4.7% of tokens, with a median span of 0.06 hours, batch deployment completed within three to four minutes; those produced by 2 to 5 creators account for 74.0% and contain 44.6% of tokens; and those produced by more than 5 creators account for only 16.8% yet contain 50.7% of tokens, with a median size of 10 tokens and a median span of 666.5 hours, roughly 28 days. Multi-creator lodes are a minority in lode count but a majority in token count, and their spans run to weeks. The bulk of name reuse is thus not instantaneous batch behavior but a name-adoption process extending over weeks, joined successively by multiple independent actors.

**Table 7 Composition of reused lodes by number of participating creators**

| Group | Lodes | Share of reused lodes (%) | Tokens | Share of tokens (%) | Median size (tokens) | Median span (hours) |
|---|---|---|---|---|---|---|
| Single creator | 6,617 | 9.2 | 17,012 | 4.7 | 2 | 0.06 |
| 2–5 creators | 52,945 | 74.0 | 162,041 | 44.6 | 2 | 0.76 |
| More than 5 creators | 12,023 | 16.8 | 184,151 | 50.7 | 10 | 666.50 |

Notes: The sample comprises the 71,585 lodes used more than once; span is the difference between the creation times of the first and last tokens within a lode. The difference between the same-creator and cross-creator interval distributions is highly significant under a one-sided Mann–Whitney test ($p < 0.001$). The median span of 0.06 hours

(about 3.6 minutes) in the single-creator group corresponds to batch deployment; the 666.5 hours (about 28 days) in the more-than-5-creators group corresponds to sustained cross-actor adoption.

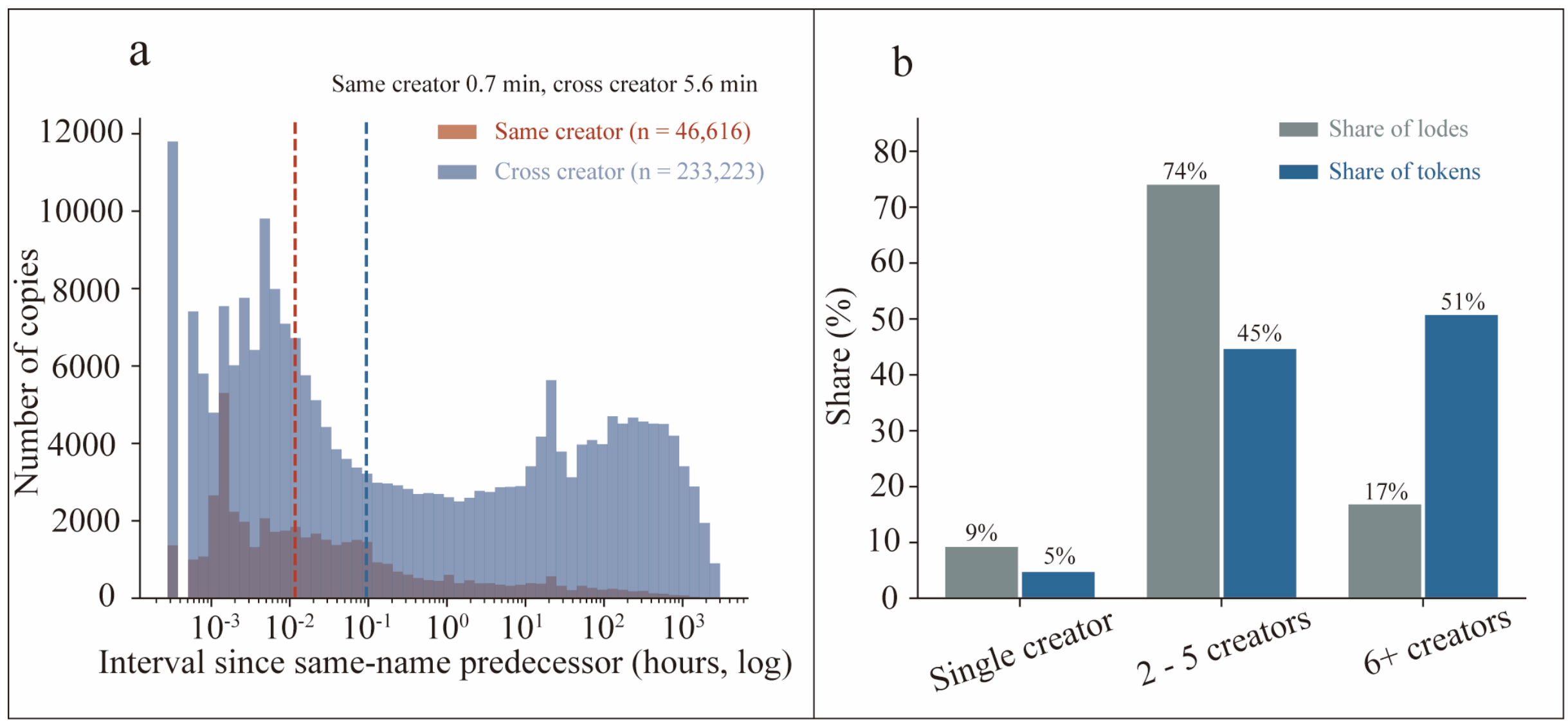

Figure 5. Decomposition of same-name copying by source. (a) Distribution of intervals to the same-name predecessor, colored by whether the copier and the predecessor's creator are the same address (logarithmic horizontal axis, dashed lines at group medians); (b) shares of lode count and token count for reused lodes grouped by number of participating creators.

### 3.5 Naming responses at event time

Cultural events leave immediate imprints on the naming population. At 12:00 UTC on February 16, 2026, 20:00 Beijing time on Lunar New Year's Eve, the CCTV Spring Festival Gala began broadcasting; within the following six hours the platform added 4,128 tokens, roughly 2.8 times the count of the six hours before the broadcast, 12.3% of them bearing festival-related names, and 136 tokens matching the string "春晚" (Spring Festival Gala) character for character (Table 8).

The hourly series (Figure 4c) shows that hourly token creation amplifies immediately after the broadcast begins, an impact measured in hours and concentrated after onset. The share of festival-related names, by contrast, shows no upward response to the broadcast: it stood at 17% to 18% throughout the twelve hours before onset, moved down rather than up in the six hours after onset (12.3%), and fell to 1.7% in the following window (Table 8). In other words, the inclination to name with festival vocabulary is set by the date, being New Year's Eve, and is elevated across the day, whereas the decision to issue at a given moment is set by the fact that the Gala is airing. The event raises the quantity of issuance and does not raise the festival content of naming.

This distinction has direct implications for event-study design. Consider a study using the broadcast moment as a cutoff to test whether the event changed naming content. The absolute number of festival-related tokens does surge after onset (total issuance amplifies by a factor of 2.8, so even at a constant share the count of festival names multiplies), and counting tokens would yield the conclusion that the event shifted naming toward festival vocabulary; the share data show that this conclusion does not hold, as the share is flat or slightly lower across the cutoff. A surge in quantity and a shift in content are two different things, the former constituting no evidence for the latter, and event studies of naming content with time cutoffs must take shares rather than counts as the outcome variable.

**Table 8  Naming responses within the Gala event window**

| Window (relative to broadcast) | Tokens created | Festival-name share (%) | Exact "春晚" matches |
|---|---|---|---|
| [−12h, −6h) | 570 | 17.5 | — |
| [−6h, 0) | 1,470 | 18.2 | — |
| [0, +6h) | 4,128 | 12.3 | 136 |
| [+6h, +12h) | 240 | 1.7 | — |

Notes: The event time is 12:00 UTC on February 16, 2026 (20:00 Beijing time on Lunar New Year's Eve, the start of the CCTV Spring Festival Gala broadcast). Festival-related names are identified by containment matching of a ten-word lexicon against the name field. Creation volume in the six hours after onset is roughly 2.8 times that of the six hours before, while the festival-name share does not rise with onset; the quantity effect and the content effect are separable in time.

## 4. Discussion

### 4.1 Portability of the framework

The components of the three-layer framework depend on the setting to different degrees, and this difference determines how the framework travels. Measurement in the form and reference layers depends only on the name string itself and a codebook, and is directly portable to any naming corpus, including app-store listing titles, domain registrations, NFT collections, short-video hashtags, and product listing names. The 38 Chinese-oriented form-layer fields apply to most Chinese naming settings and fill the gap that fluency measurement has left on the Chinese writing system. The economic significance of that gap extends beyond crypto assets: Chinese brand naming, corporate renaming in A-share markets, and ranking competition in Chinese app stores have all lacked a computable fluency yardstick. Porting the relation layer requires two institutional conditions, namely that names may be reused (no uniqueness constraint, or a weak one) and that creation timestamps are available. In settings with enforced uniqueness, such as stock tickers and domain names, lodes and naming order do not exist in principle, but the other two feature families remain constructible: crowding requires only the set of contemporaneous names, and lexical variation only a historical name repository.

### 4.2 Research potential of the data

The census dataset delivered with this paper contains research possibilities the paper itself does not use. No prior study has reported, for a zero-cost naming market, the distribution of name reuse, the speed of copying, or the shape of event responses; these population quantities were previously unavailable not because they are unimportant but because outcome-sampled data cannot produce them.

For linguistics and cultural evolution, the Heaps exponent of the name space, the heavy-tail exponents, and the decay path of the new-name share provide a rare testing ground for models of lexical innovation and cultural transmission [38,39]: in natural-language corpora the birth and each adoption of a word cannot be observed item by item, whereas this dataset records every naming event completely with second-level timestamps, and the near-zero cost of creation compresses supply-side friction to a minimum.

For research on imitation and diffusion, the two-scale structure of copying (0.7 minutes within creator versus 5.6 minutes across creators) and the weeks-long adoption process of multi-creator lodes provide contagion observations with identifiable actors and precise timing; the record of who adopted whose innovation at what moment, which social-contagion and innovation-diffusion models require, is here traceable item by item.

For market design, the crowding rate documented here (the new-name share declining by roughly three percentage points per month) and the concentration pattern (the largest 1% of lodes containing 17.4% of tokens) characterize how a name space without property-rights constraints organizes itself, and serve as empirical input to the design question of whether names should be subject to uniqueness or property-rights constraints.

### 4.3 Limitations

The first limitation of this paper concerns scope: it does not test the relation between naming and market outcomes, and its validity evidence is confined to convergent validity (crowding rises monotonically with concurrency, and lexical variation agrees in direction with manual homophone coding), discriminant validity (the three layers are nearly orthogonal), and structural validity (the co-occurrence lifts of the six dimensions accord with semantic relations); predictive validity is not among them and awaits the companion study. In addition, the six coding dimensions are tailored to this setting, and porting them to other settings may require rebuilding the codebook; what transfers is the triple-benchmark reliability protocol itself rather than the content of the dimensions. The reproducibility of LLM coding is constrained by model availability: prompts and model versions are archived with the data package, but the long-term availability of the model is beyond the authors' control. For this reason the 408 double-blind human annotations are released with the paper, and any subsequent model can be recalibrated against them. Finally, the descriptive regularities derive from a single platform and a single time window; whether the Heaps exponent, the tail exponents, the copying intervals, and the magnitude of event responses are stable across platforms and periods awaits parallel censuses.

## 5. Conclusion

This paper builds a census-level dataset of all 513,647 naming attempts over five months on the four.meme token launchpad on BNB Chain and proposes and validates a three-layer measurement framework for names. The form layer replaces inapplicable letter-sequence indicators with 38 deterministic fields adapted to the Chinese writing system, extending fluency measurement to Chinese-named assets. The reference layer combines a six-dimension codebook, double-blind two-round human coding, and full-scale LLM expansion, with reliability reported on the three benchmarks of human double-blind agreement, human–machine replication, and machine self-consistency. The relation layer measures, for the first time systematically, the position of a name within the population of names. The three layers are nearly orthogonal in the data, each carrying independent information.

The census reveals four population regularities previously unobservable. The name space expands according to Heaps' law ($V = 1.24 \cdot N^{0.921}$, $R^2 = 0.999$), the share of new names declines monotonically from 52.0% to 37.8%, and the language composition drifts by 13.5 percentage points over the same period. Name reuse is heavy-tailed (tail exponents 2.19 to 2.67), with the largest 1% of lodes containing 17.4% of tokens. The median interval between successive uses of the same name is under three minutes and decomposes by creator identity into two separable time scales: batch deployment at 0.7 minutes within creator, and imitative adoption across creators starting at 5.6 minutes and extending over weeks. Cultural events trigger immediate quantity responses, while the use of festival vocabulary follows the calendar rather than the moment, the two responses being separable in time.

The significance of these results is not confined to the crypto-asset setting. Methodologically, the paper delivers an infrastructure that renders names computable variables, comprising the three-layer framework, a fluency yardstick for the Chinese writing system, and a reliability protocol reusable in any LLM-assisted content analysis. In terms of data, a naming population that is completely sampled, nearly costless to enter, and identifiable by actor and moment provides a previously nonexistent object of observation for research on lexical innovation, imitation and diffusion, and the market design of name property rights.

## CRediT authorship contribution statement

Dingding Cao: Conceptualization, Methodology, Software, Data curation, Formal analysis, Writing – original draft. Yujing Zhong: Investigation, Validation, Data curation. Han Wang: Investigation, Validation. Xian Pan: Resources, Visualization. Wei Yang and Rizwan Akhtar: Supervision, Project administration, Writing – review and editing.

**Funding**

This work was supported by the Baoshan University Doctoral Research Startup Fund Program 2025 (Grant No. BSKY2542) and the Youth Project of the Local Universities Joint Fund of Yunnan Province Science and Technology Plan (Grant No. 202101BA070001-272).

**Declaration of competing interest**

The authors declare no competing interests. The authors hold no positions in any token in the sample.

**Data and code availability**

All replication materials are archived on Zenodo (https://doi.org/10.5281/zenodo.22177782): the census master table; the 408 double-blind annotations and the codebook (frozen Chinese original with an English translation); the implementation code for all three measurement layers and the scripts producing every table, figure, and in-text statistic; and the reference table for the 4,819 Chinese characters in the corpus. Raw trade-level data (approximately 15 GiB) are available under controlled access. Every number reported in this paper can be independently recomputed from these materials.

**Declaration on the use of large language models**

A large language model was used solely for the full-scale coding expansion of the six judgment dimensions in the reference layer. The model is DeepSeek-V4-Flash-0731 (API alias deepseek-v4-flash); the calling parameters (temperature 0, fixed seed, batch size), the full prompts and their hashes, and the actual model versions returned by the server batch by batch are archived with the data package. Beyond this, large language models were used only for language polishing and for the translation and refinement of code. All text of the paper was written by the authors, who bear full academic responsibility.

**Ethics statement**

This research uses only public on-chain data and public APIs. On-chain data require no anonymization.